\documentstyle[epsfig,twocolumn,amsfonts,prl,aps]{revtex}
\begin{document}
\draft
\title{Charge Transfer in Cluster--Atom Collisions\\
        Studied with Non-Adiabatic Quantum Molecular Dynamics}
\author{O. Knospe\cite{perm}, J. Jellinek}
\address{Chemistry Division, Argonne National Laboratory, 
         Argonne, Illinois 60439, USA}
\author{U. Saalmann}
\address{Max-Planck-Institut f\"ur Physik komplexer Systeme,
         N\"othnitzer Str.~38, 01187 Dresden, Germany}
\author{R. Schmidt}
\address{Technische Universit\"at Dresden, Institut f\"ur 
         Theoretische Physik, 01062 Dresden, Germany}
%\date{\today}
\maketitle
\begin{abstract}
Charge transfer in collisions of Na$_n^+$ cluster ions with Cs 
atoms is investigated theoretically in the microscopic framework
of non-adiabatic quantum molecular dynamics. The competing 
reaction channels and related processes affecting the charge 
transfer (electronic excitations, fragmentation, temperature) 
are described. Absolute charge transfer cross sections for 
Na$_n^+$(2.7 keV) + Cs  
$\longrightarrow {\rm Na}_n \: + \: {\rm Cs}^+$
have been calculated in the size range $4\le n \le 11$ reproducing 
the size dependence of the experimental cross sections. The energy
dependence of the cross section is predicted for $n=4,7,9$. An 
exotic charge transfer channel producing Cs$^-$ is found to have 
a finite probability.
\end{abstract}

%\narrowtext
Charge transfer represents one of the fundamental atomic 
interactions. In the last decades, large progress has been made in 
the understanding of charge transfer mainly for two borderline cases: 
the elementary ion--atom reaction \cite{BMcD} and the complex 
ion--surface interaction \cite{surf}. In the recent past, however, there 
has been a flurry of activity to close the gap between the two fields 
by investigating the intermediate case of ion--cluster collisions 
\cite{Bre1,Bre2,l7-11,l7-7,l8-29,l8-23,l8-12,l10-1,l8-14,ulf-23,l8-19,l8-18,l8-28,l8-21,Sidis,Borre}.
In general, the basic aspect of cluster collisions consists in the 
simultaneous occurrence and mutual coupling of electronic 
transitions (charge transfer, excitation, ionization) and nuclear-core 
excitations (vibration, rotation, fragmentation) in a system, where the 
number of electronic and nuclear degrees of freedom is {\it large but 
finite}. In particular, collision induced dissociation (CID) will compete 
with charge transfer (CT) in such collisions.

Already, in one of the first experiments with mass-selected cluster 
beams, CT and CID were investigated for Na$_n^+$ + Cs  
\cite{Bre1,Bre2} and K$_n^+$ + Cs \cite{Bre2} collisions. An 
important result of these novel experiments was the direct 
determination of the CT cross section for the neutralization of the 
parent cluster ion, i.e.
\begin{equation}\label{eq1} 
 {\rm Na}_n^+ + {\rm Cs} \longrightarrow {\rm Na}_n + {\rm Cs}^+ 
\end{equation}
by measuring the neutral products associated with the process 
(\ref{eq1}) but discriminating neutral products originating from CID. 
Further experimental studies of CT in collisions of cluster cations 
with atoms \cite{l7-11,l7-7,l8-29,l8-23,l8-12,l10-1}, 
molecules \cite{l8-29} and clusters \cite{l8-14,l8-12} have been 
performed providing a considerable amount of data. Multiple CT 
processes accompanied by fragmentation have been observed in 
collisions of highly charged ions with fullerenes \cite{ulf-23,l8-19} 
and sodium clusters \cite{l8-18}.

This enormous experimental progress was accompanied by the 
development of a number of theoretical descriptions of CT reactions 
in cluster collisions. In Ref.~\cite{Bre2}, a two-state model of (near) 
resonant CT was presented. Classical barrier models 
\cite{l8-28,l8-21} have been applied to distant collisions of C$_{60}$ 
with highly charged ions. Semi-microscopic descriptions of CT, 
which are based on the jellium approximation \cite{Sidis} or on 
phenomenological single-particle potentials \cite{Borre}, have also 
been suggested. However, the combined description of CT {\it and} 
fragmentation or, more generally, a simultaneous treatment of 
electronic {\it and} nuclear degrees of freedom in non-adiabatic 
cluster collisions is still an unsolved problem. 

Recently, a universal microscopic approach called non-adiabatic 
quantum molecular dynamics (NA-QMD) has been developed, which 
describes classical atomic motion simultaneously and 
self-consistently with electronic transitions in atomic many-body 
systems \cite{ulf}. In this paper, we present a fully microscopic 
analysis of CT and fragmentation in cluster collisions applying the 
NA-QMD theory to Na$_n^+$ + Cs.  

The NA-QMD approach has been derived \cite{ulf} on the basis of 
the time-dependent density functional theory \cite{Gross}, whereby 
the Kohn--Sham formalism within the time-dependent local density 
approximation is used. Resulting from an LCAO ansatz for the 
Kohn--Sham orbitals a set of coupled differential equations for the 
time-dependent coefficients is obtained to determine the time 
evolution of the electronic density as the consequence of the 
classical atomic motion. Simultaneously, Newton's equations of 
motion with explicitly time-dependent forces have to be solved 
reflecting the possible energy transfer between the classical system 
of ionic cores and the quantum-mechanical system of valence 
electrons. Details of the procedure to calculate
expectation values of observables or probabilities for specific
electronic transitions (defined by the time-dependent electronic 
density) as well as of the different fragmentation channels (obtained 
from classical trajectories) will be given elsewhere \cite{CT2}.

\begin{figure}
\centerline{\hbox{\psfig{figure=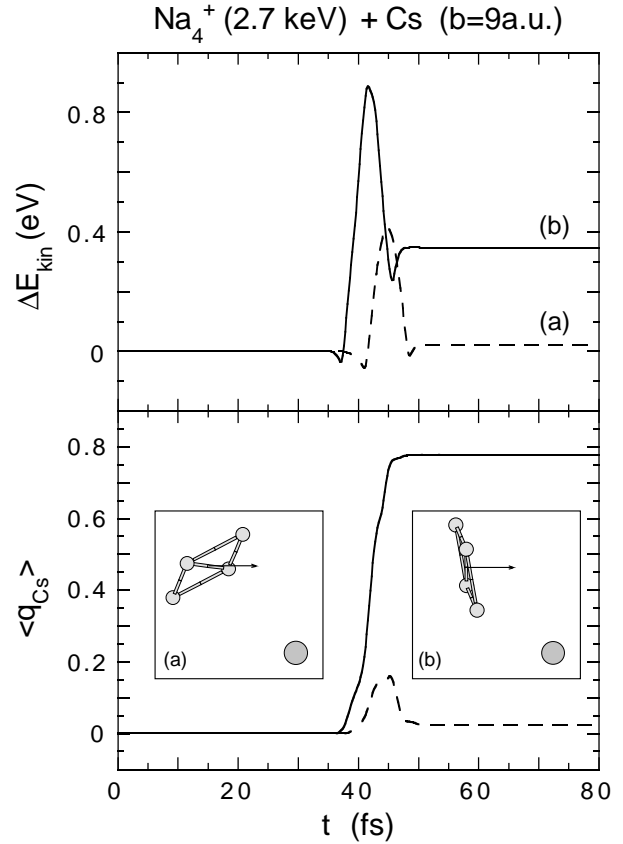,width=8.4cm}}}
\caption{Calculated kinetic-energy difference $\Delta E_{\rm kin}$ 
(upper panel) and mean value of the 
charge located at the Cs atom $\langle q_{\rm Cs}\rangle$ 
(lower panel) as functions of time $t$ for the two initial collision 
geometries illustrated in the inserts (a) and (b). The dashed and 
solid lines correspond to the cases (a) and (b), respectively.}
\label{fig1}
\end{figure}

To elucidate the different physical processes determining the 
absolute cross section of the CT reaction (\ref{eq1}), we have 
performed a detailed study of the system Na$_4^+$ + Cs at a lab 
collision energy of $E_{\rm lab} =2.7 \:{\rm keV}$. The cluster 
projectile is prepared initially in its electronic and geometric ground 
state (rhombic, $D_{\rm 2h}$). In Fig.~\ref{fig1}, the time evolution of 
the calculated mean charge located at the Cs atom $\langle q_{\rm 
Cs}\rangle(t)$ (obtained from a population analysis) as well as of the 
kinetic-energy difference 
$\Delta E_{\rm kin}(t)=E_{\rm c.m.}-E_{\rm kin}(t)$, where 
$E_{\rm c.m.}$ and $E_{\rm kin}(t)$ are the collision energy and the 
total kinetic energy, respectively, both referred to the center-of-mass 
frame, are presented for two collision events with the same impact 
parameter ($b=9\:{\rm a.u.}$) but different initial orientations of the 
cluster with respect to the beam axis (see inserts (a) and (b) in
Fig.~\ref{fig1}). The quantity $\langle q_{\rm Cs}\rangle(t)$ 
characterizes the quantum dynamics of CT, whereas 
$\Delta E_{\rm kin}(t)$ describes the energy flow in the classical 
(ionic) degrees of freedom. A strong dependence of both quantities 
on the initial orientation of the cluster is observed. In the example 
(a), nearly compensating charge fluctuations during the interaction 
lead finally to $\langle q_{\rm Cs}\rangle \approx 0$, and the event 
corresponds to a nearly elastic scattering $\Delta E_{\rm kin} \approx 
0$. In contrast, the large CT in the case (b) 
$\langle q_{\rm Cs}\rangle \approx 0.75$ is accompanied by a 
considerable energy loss of $\Delta E_{\rm kin} \approx 0.34\:{\rm 
eV}$. An endothermic character is typical for most collision events 
with appreciable CT, even at larger impact parameters, where no 
vibrational excitation of the cluster occurs. Comparing the ionization 
potentials of the Cs atom (3.89 eV) and the Na$_4$ cluster (4.24 eV 
\cite{Kap}) and assuming a CT that results in the electronic ground 
state of Na$_4$, one should expect an exothermic reaction with 
$\Delta E_{\rm kin} \approx - 0.35\:{\rm eV}$. That this is not the 
case indicates that the CT process produces an electronically 
excited Na$_4$ cluster. 

\begin{figure}
\centerline{\hbox{\psfig{figure=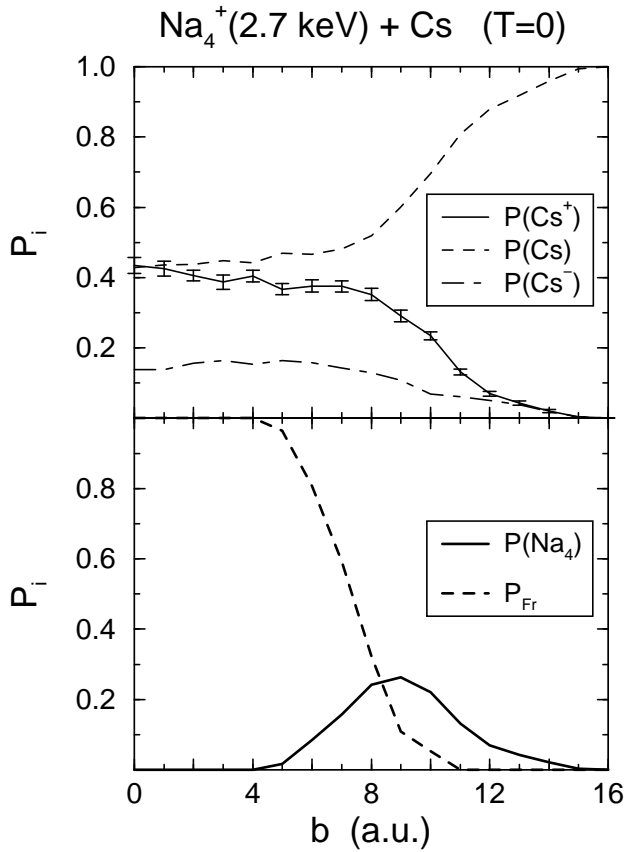,width=8.4cm}}}
\caption{Calculated integral CT probabilities $P({\rm Cs}^+)$, 
$P({\rm Cs})$, $P({\rm Cs}^-)$ [upper panel], and fragmentation 
$P_{\rm Fr}$ and CT probabilities $P({\rm Na}_4)$ [lower panel] as 
functions of the impact parameter $b$ for Na$_4^+$ chosen initially 
at $T=0\:{\rm K}$ in its lowest-energy structure (see text). The length 
of the error bars is given by $2\,s_{\rm D}$, where $s_{\rm D}$ is the 
standard deviation of the orientation average.} 
\label{fig2}
\end{figure}

The calculation of CT cross sections requires a detailed analysis of 
the final electronic and atomic states and, in particular, the careful 
consideration of CT probabilities for different reaction channels. To 
start with a transparent classification, we define integral CT 
probabilities, which are the probabilities $P({\rm Cs}^q)$ to find the 
Cs atom in the charge state $q$ in the exit channel, where $q$ is an 
integer and $\sum_q P({\rm Cs}^q)=1$. These probabilities describe 
the primary CT ($q\neq 0$) and scattering ($q=0$) processes 
without regard to the further evolution of the cluster, i.e.~to the 
possible fragmentation. The related CT cross sections can be 
directly measured by detecting the formed Cs ions. The probabilities 
$P({\rm Cs}^q)$, calculated as an average of the results obtained 
with about 300 different initial orientations of the cluster per impact 
parameter $b$, are shown as a function of $b$ in the upper panel of 
Fig.~\ref{fig2}. The CT leading to Cs$^+$ ions and the scattering 
without CT have nearly equal probabilities $P({\rm Cs}^+)$ and 
$P({\rm Cs})$, respectively, for impact parameters $b\le 8\:{\rm 
a.u.}$. As the graph of $P({\rm Cs}^+)$ indicates, CT takes place 
with remarkable probability up to impact parameters of 
$b\approx 15\:{\rm a.u.}$, which is more than twice the long half-axis 
($R=5.7\:{\rm a.u.}$) of the Na$_4$ rhombus. Surprisingly, the 
calculations also yield a finite probability $P({\rm Cs}^-)$ for an 
electron transfer to the Cs atom, which represents an interesting 
prediction for future experimental studies. 

The $b$-weighted integration of $P({\rm Cs}^+)$ leads to a total CT 
cross section of $\sigma({\rm Cs}^+)=(38.2 \pm 2.1)\:{\rm \AA}^2$. 
For comparison, the ``geometrical'' cross section of the cluster is 
about $\sigma_0 \approx \pi\,R^2 \approx 29 \:{\rm \AA}^2$ and the 
measured value is 
$\sigma^{\rm exp}({\rm Na}_4) = (17\pm 3) \: {\rm \AA}^2$ 
\cite{Bre1}. It is important to realize, however, that, instead of 
Cs$^+$, the signal of the neutralized Na$_4$ was detected in the 
measurements \cite{Bre1}. The ``survival'' probability of the Na$_4$ 
cluster after the primary CT is given by 
$P({\rm Cs}^+)\cdot \left[ 1 - P_{\rm Fr} \right]$, where $P_{\rm Fr}$ 
is the total fragmentation probability. In cluster collisions,
fragmentation can be induced through three mechanisms 
characterized by different time scales \cite{CT2,Barat1,Barat2}. 
These are large momentum transfer between atoms of the projectile 
and of the target, electronic excitation followed by energy transfer 
via electron-vibrational coupling, and statistical fragmentation. 
Whereas the first two mechanisms are precisely described by the 
NA-QMD theory and automatically accounted for in the actual 
calculations (this has been shown in comparisons with experimental 
data on CID \cite{Barat2}), the total probabilities of the different 
fragmentation channels can be determined from the calculated 
internal (vibrational and electronic) excitation energy of the cluster 
using statistical arguments \cite{CT2}. The calculated total 
fragmentation probability $P_{\rm Fr}$ and the probability $P({\rm 
Na}_4)$ to have a neutral Na$_4$ cluster in the exit channel are 
shown as a function of $b$ in the lower panel of Fig.~\ref{fig2}. 
\noindent
Fragmentation occurs with high and then decreasing probability up 
to $b\approx 11\:{\rm a.u.}$, and the computed total fragmentation 
cross section of $\sigma_{\rm Fr} = 49.4\:{\rm \AA}^2$ exceeds the 
geometrical cross section considerably. The CT probability 
$P({\rm Na}_4)$ peaks around $b\approx 9\:{\rm a.u.}$, which is 
larger than the cluster ``radius'' $R$. The CT cross section 
$\sigma({\rm Na}_4) = 20.2 \: {\rm \AA}^2$ is still slightly larger
than the experimental value. 

The calculations presented so far have been carried out with zero 
initial temperature of the cluster, whereas under the experimental 
conditions \cite{Bre1,Bre2} ``liquid'' cluster ions are used resulting 
from the laser ionization to produce the cluster beam 
\cite{Haberland}. To take into account the temperature effect, we 
have repeated the procedure described above with excited clusters 
in the initial state, where the total excitation energy was chosen 
to be slightly below the dissociation energies of the neutral as 
well as of the cationic cluster. The initial configurations of the excited 
clusters for the simulation of the collisions were produced in 
equilibration runs of 300 ps in length. The calculated CT cross 
section $\sigma({\rm Na}_4) = 16.8 \: {\rm \AA}^2$ obtained with 
``liquid'' cluster ions is in perfect agreement with the experimental 
result. 

\begin{figure}
\centerline{\hbox{\psfig{figure=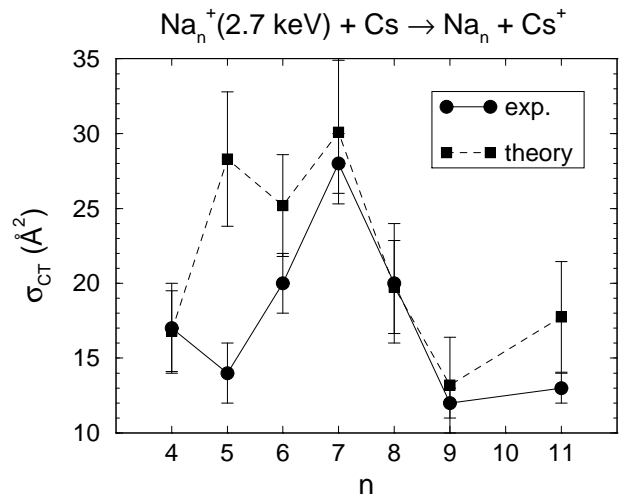,width=8.4cm}}}
\caption{Calculated and measured \protect\cite{Bre1} CT cross 
sections $\sigma_{\rm CT}({\rm Na}_n)$. The theoretical error bars 
result from the statistical uncertainty $2\,s_{\rm D}$ 
(cf.~Fig.~\protect\ref{fig2}).}
\label{fig3}
\end{figure}

The theoretical and experimental \cite{Bre1} CT cross sections are 
compared as a function of the cluster size $n$ in Fig.~\ref{fig3}. The 
theoretical results were obtained with ``liquid'' cluster ions in the 
initial state. Except for the case of $n=5$, the agreement of the 
computed and measured data can be qualified as perfect, since 
these data are {\it absolute} cross sections. The statistical 
uncertainties of the calculated cross sections reflect the strong 
dependence of the outcome of the collision process on the initial 
configuration. These uncertainties could be reduced further only at a 
very high computational cost \cite{CT2}. The overestimated cross 
section in the case $n=5$ results from contributions of a particular 
isomer, for which accidentally a (quasi-) resonant CT with Cs occurs 
\cite{CT2}.

\begin{figure}
\centerline{\hbox{\psfig{figure=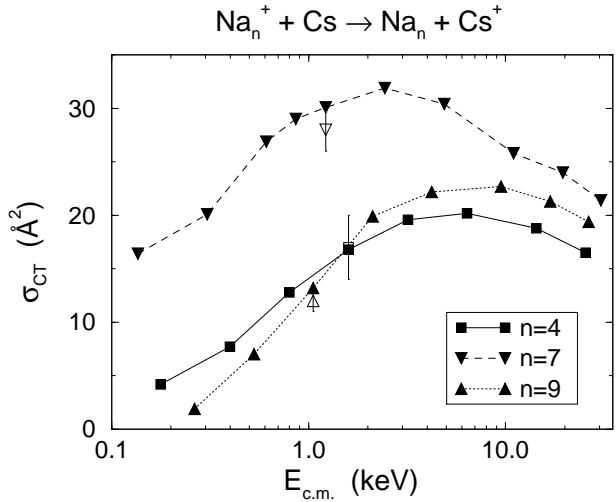,width=8.4cm}}}
\caption{Calculated collision energy dependence of the CT cross 
section $\sigma_{\rm CT}({\rm Na}_n)$ for $n=4,7,9$ (full symbols). 
The experimental data (open symbols with error bars; 
\protect\cite{Bre1}) correspond to the same lab collision energy of 
$E_{\rm lab} = 2.7 \: {\rm keV}$.} 
\label{fig4}
\end{figure}

The distinct maximum of the experimental CT cross section at $n=7$ 
is reproduced by the calculations. This maximum, however, may be 
peculiar to the particular collision energy. In the experimental 
investigation of K$_n^+$ + Cs collisions \cite{Bre2} a very strong 
dependence of the CT cross section on the collision energy has 
been found (cf.~Fig.~4 in Ref.~\cite{Bre2}). In order to examine this 
aspect and to stimulate further experimental investigations, we have 
calculated the CT cross section over a wide range of the 
(center-of-mass) collision energy 
($0.1\:\ldots\: 30\:{\rm keV}$) for $n=4,7,9$. The results are shown in 
Fig.~\ref{fig4}. The absolute CT cross section for $n=7$ exceeds 
those for $n=4$ and $n=9$ over the entire energy range considered. 
Consequently, the large CT cross section in Na$_7^+$ + Cs 
collisions should be attributed to the specific electronic structure of 
Na$_7^+$ providing favorable, i.e.~near resonant conditions for CT 
in collisions with Cs. 

In summary, we have presented results of a fully microscopic 
analysis of CT and fragmentation in cluster-ion -- atom collisions 
based on NA-QMD simulations. The detailed study of 
Na$_4^+$(2.7 keV) + Cs collisions revealed the role of the different 
physical processes associated with CT in cluster collisions 
(electronic excitations, fragmentation, temperature effects). An exotic 
``inverse'' CT process leading to Cs$^-$ is predicted. The calculated 
absolute CT cross sections for 
Na$_n^+$(2.7 keV) + Cs ($4\le n \le 11$) 
are in good agreement with the experimental data. The energy 
dependence of the CT cross section is predicted for several cluster 
sizes ($n=4,7,9$) in order to encourage further experimental studies. 

This work was supported by the Office of Basic Energy Sciences, 
Division of Chemical Sciences, US DOE under contract 
W-31-109-ENG-38 (O.K., J.J.), by the DFG through 
Schwer\-punkt ``Zeitabh\"angige Ph\"anomene und Methoden in 
Quantensystemen in der Physik und Chemie'' (O.K., U.S., R.S.), and 
by the EU through the HCM network CMRX-CT94-0614 (O.K., R.S.). 
One of us (O.K.) acknowledges gratefully a scholarship from the 
DAAD.

\end{document}